\documentclass[ams, prb, twocolumn, amssymb,floatfix]{revtex4-1}
\usepackage{amsmath}
\usepackage{tabularx}
\usepackage{bm}
\usepackage{euscript}
\usepackage{graphicx}
\usepackage{color}
\usepackage{amsfonts}
\usepackage{exscale}
\usepackage{amsbsy}
\usepackage{subfigure}
\usepackage{textcomp}
\usepackage{comment}

\usepackage{hyperref}

\newcommand{\beq}{\begin{equation}}
\newcommand{\eeq}{\end{equation}}
\newcommand{\be}{\begin{eqnarray}}
\newcommand{\ee}{\end{eqnarray}}

\begin{document}

\title{Composite fermion Hall conductivity and the half-filled Landau level}
\author{Prashant Kumar,$^1$ S. Raghu,$^{1,2}$ and Michael Mulligan$^{3}$}
\affiliation{$^{1}$Stanford Institute for Theoretical Physics, Stanford University, Stanford, California 94305, USA}
\affiliation{$^2$SLAC National Accelerator Laboratory, 2575 Sand Hill Road, Menlo Park, CA 94025, USA}
\affiliation{$^3$Department of Physics and Astronomy, University of California, Riverside, CA 92511, USA}
\date{\today}

\begin{abstract}
We consider the Hall conductivity of composite fermions in the theory of Halperin, Lee, and Read (HLR).
We present a fully quantum mechanical numerical calculation that shows, under suitable conditions, the HLR theory exhibits a particle-hole symmetric dc electrical Hall response in the presence of quenched disorder.
Remarkably, this response of the HLR theory remains robust even when the disorder range is of the order of the Fermi wavelength.
We find that deviations from particle-hole symmetric response can appear in the ac Hall conductivity at frequencies sufficiently large compared to the inverse system size.
Our results agree with a recent semi-classical analysis by Wang {\it et al.}, Phys.~Rev.~X 7, 031029 (2017) and complement the arguments based on the fully quantum-mechanical model by Kumar {\it et al.}, Phys. Rev. B 98, 11505 (2018).
These results provide further evidence that the HLR theory is compatible with an emergent particle-hole symmetry.
\end{abstract}

\maketitle

\section{Introduction}

In the limit of an infinitely strong perpendicular magnetic field, electrons in two spatial dimensions (2D) are spin polarized and their dynamics is governed entirely by the lowest Landau level (LLL). 
When the LLL is half-filled ($\nu = 1/2$), the electron fluid can exhibit Landau level particle-hole symmetry: the electrical Hall conductivity must equal $\sigma_{xy} = 1/2 $ in units\footnote{In the remainder of the paper, we set $e = \hbar = 1$ making the quantum of conductance $\frac{e^2}{h} = \frac{1}{2 \pi}$.} of $e^2/h$, as a consequence.\cite{kivelson1997}
This result holds for frequencies sufficiently small compared with the cyclotron energy.
Particle-hole symmetry therefore places an important constraint on any long wavelength description of the half-filled Landau level.\footnote{Particle-hole symmetry in the LLL is broken by a quenched random chemical potential.  
However, disorder-averaged conductivities can be particle-hole symmetric, if the disorder has vanishing moments $\overline{V({\bf x}_1) \cdots V({\bf x}_m)} = 0$ for all odd $m$.}

A useful description of the dynamics of electrons at $\nu=1/2$ involves composite fermions.
\cite{Jainbook, Fradkinbook}
There are two prominent composite fermion effective field theories.  
The first, pioneered by Halperin, Lee and Read (HLR),\cite{halperinleeread} involves non-relativistic composite fermions, which can be viewed as electrons bound to two flux quanta.  
Because the particle-hole transformation does not act in a transparent fashion on the HLR Lagrangian\cite{kivelson1997, BMF2015}, it has been unclear whether the HLR theory is compatible with particle-hole symmetry, {\it i.e.}, if there is a limit of the HLR theory in which the theory is particle-hole symmetric.
By contrast, an alternate theory,\cite{Son2015} recently constructed by Son, manifestly preserves particle-hole symmetry.  
The composite fermions in Son's theory are electrically neutral fermionic vortices of the electron fluid and are described by a Dirac Lagrangian.   
  
Despite the differences in their formulations, recent work suggests that the composite fermion theories of HLR and Son may have the same long wavelength experimental consequences.
In an insightful recent study by Wang {\it et al.},\cite{2017PhRvX...7c1029W} a semi-classical analysis demonstrated that a number of observables in the HLR theory can indeed exhibit particle-hole symmetry if the effects of quenched disorder are properly incorporated. 
We have recently studied the HLR mean-field theory with disorder beyond the semi-classical approximation.\cite{PhysRevB.98.115105} In particular, we found that the particle-hole symmetry in dc Hall response of the HLR theory is a result of the supersymmetric quantum-mechanical structure of the composite fermion Hamiltonian. 
This relation, which relies on the precise correlation (see Eq.~\eqref{slaved}) that flux attachment imposes between the potential and magnetic field disorders that the composite fermions experience, enabled us to argue that the HLR theory represents a quantum phase transition between an insulator and an integer quantum Hall state of composite fermions. 
Universality of the resistivity tensor\cite{Shahar1995} at such a phase transition leads to particle-hole symmetric Hall conductivity.
In addition, within a mean-field approximation, the HLR and Son theories have been found to produce coincident quantum oscillation minima in the presence of an external periodic potential.\cite{PhysRevB.95.235424}

The implication of these works is that particle-hole symmetry in the HLR theory emerges at low energies as a property of an IR fixed point of the renormalization group, rather than being manifest in the bare Lagrangian.\footnote{See \cite{PotterSerbynVishwanath2015, PhysRevB.94.245107, BalramJain2016, LevinSon2016} for other studies contrasting the HLR and Dirac composite fermion theories and earlier numerical work\cite{rezayi2000, Geraedtsetal2015} that found evidence for a particle-hole symmetric electron ground state at $\nu=1/2$.}

However, there are further questions to which numerical calculations can provide answers. First, in the work of Wang {\it et al.}, the dc Hall conductivity was obtained from a $1/{\left( k_F \cal R \right)}$ expansion, valid when the range $\cal R$ of disorder is long compared to the Fermi wavelength $2\pi/k_F$.
Is particle-hole symmetry preserved at higher orders in this expansion?
Second, does the Hall conductivity maintain its particle-hole symmetric value at non-zero frequencies?
Third, is supersymmetry of the composite fermion Hamiltonian necessary for particle-hole symmetric dc Hall response? 

Our goal here is to answer these three questions.
To this end, we present fully quantum mechanical calculations of the composite fermion Hall conductivity in the presence of quenched disorder.
We study the problem numerically using a continuum model and confirm that the particle-hole symmetric dc response can survive in the HLR theory even for disorder configurations that vary on the order of the Fermi wavelength (see Fig.~\ref{continuum_plots}).
This is in agreement with the conclusions of Ref. \onlinecite{PhysRevB.98.115105}.
At non-zero frequencies, we observe a deviation from particle-hole symmetric Hall response (see Fig.~\ref{hall_vs_L}), the size of which is reduced as the system size of our model is increased. Lastly, we study the degree to which the particle-hole symmetric response is robust against perturbations that violate the supersymmetric structure of the HLR Hamiltonian: we study composite fermions on a lattice with nearest neighbor hopping in the presence of disorder (see Fig. \ref{lattice_plots}).

\section{Particle-Hole symmetry and composite fermions: brief review}
  
Electrons in a half-filled LLL are described by a Lagrangian of the form,
\begin{equation}
 \mathcal L_{\rm el}= \psi^{\dagger}(r)  \left( i \partial_t + \mu +  A_t  -  \frac{1}{2 m} \left( i \partial_j + A_j \right)^2\right) \psi(r)  + \mathcal L_{\rm int},
 \end{equation}
where $\psi(r)$ destroys a spin-polarized electron of mass $m$ at position $r = (t, {\bf x})$, $A_t({\bf x})$ is the electromagnetic scalar potential, $\bm A({\bf x})$ is the electromagnetic vector potential corresponding to the transverse magnetic field $B_{\perp}(\bm x) = \nabla \times \bm A({\bf x}) $, $\mu$ is the chemical potential adjusted so that the Landau level is half-filled, $ B_{\perp}(\bm x) = 4 \pi \langle \psi^{\dagger} \psi(\bm x)\rangle $, and $\mathcal L_{\rm int}$ represents interactions involving pairs of electrons ({\it e.g.}, the Coulomb interaction).  
In the composite fermion theory of HLR,\cite{halperinleeread} the low-energy behavior is postulated to be governed by a new set of fermions with Lagrangian, 
\begin{align}
\mathcal L_{\text{eff}} = \mathcal L_f + \mathcal L_{\rm cs} + \mathcal L_{\rm int},
\end{align}
where
\begin{align}
\mathcal L_f  & = f^{\dagger} \left( i \partial_t  + \mu +  A_t + a_t - \frac{1}{2 m_f} \left( i \partial_j + A_j + a_j \right)^2 \right) f, \cr
\mathcal L_{\rm cs} & = \frac{1}{2} \frac{1}{4 \pi} \epsilon_{\mu \nu \lambda} a_{\mu} \partial_{\nu} a_{\lambda}, \cr
\mathcal L_{\rm int} & = - \int d^2 x' f^{\dagger}(t, \bm x) f(t, \bm x) U (\bm x - \bm x') f^{\dagger}(t, \bm x')f(t, \bm x').
\end{align}
In ${\cal L}_{\rm eff}$, $f(r)$ destroys a composite fermion of effective mass $m_f$ at position $r$, $a_{\mu}$ is an emergent $U(1)$ gauge field, $U({\bf x})$ is a two-particle interaction, and the anti-symmetric symbol $\epsilon_{t x y} = 1$. 
The Chern-Simons term, $\mathcal L_{\rm cs}$, implements the attachment of two units of emergent gauge flux to every composite fermion.
This follows from the equation of motion of $a_t$ that sets $4 \pi \langle f^{\dagger} f(\bm x) \rangle = - \epsilon_{ij} \langle \partial_i a_j (\bm x) \rangle$.  
Furthermore, since the electron and composite fermion densities are equal, $\langle \psi^{\dagger} \psi({\bf x}) \rangle = \langle f^{\dagger} f({\bf x})\rangle$, composite fermions feel vanishing effective magnetic field on average, $B_{\rm eff} = B_{\perp}(\bm x) + \epsilon_{ij} \langle \partial_i a_j (\bm x) \rangle \approx 0$.   
Ignoring the fluctuations of $a_{\mu}$, the mean-field ground state of the composite fermions is a filled Fermi sea.

To determine the condition that composite fermions must satisfy in order for the electrons to exhibit particle-hole symmetric electrical response,  we can make use of an exact relation between the electrical and the composite fermion conductivity tensors.  
This relation can be deduced without explicitly taking quenched disorder into account.
Intuitively, since composite fermions carry two flux quanta, their current generates an additional electrical Hall voltage which in turn alters their Hall resistivity (see Appendix).  
More formally, the relation between conductivities is obtained by integrating out the composite fermions and emergent gauge field $a_{\mu}$ to obtain an effective response theory for the external field $A_{\mu}$ (see Appendix).  
As long as the composite fermion longitudinal conductivity is non-zero, particle-hole symmetric electrical response ($\sigma_{xy} = 1/4\pi$) implies that the composite fermion Hall conductivity,\cite{kivelson1997}
\begin{align}
\label{phconstraint}
\sigma_{xy}^{\rm cf} = - {1 \over 4 \pi}.
\end{align}
Thus, the composite fermion metal must have a large Hall response in order to satisfy particle-hole symmetry, an  unseemly request for a metal in vanishing effective magnetic field.  

However, as discussed by Wang {\it et al.},\cite{2017PhRvX...7c1029W} this naive expectation of vanishing Hall response is incorrect.
Since the Chern-Simons Lagrangian breaks time-reversal symmetry, a Hall response for the composite fermions is not forbidden.
Although, the large value required of $\sigma_{xy}^{\rm cf}$ may be surprising.
Because the analysis leading to Eq.~\eqref{phconstraint} requires finite $\sigma_{xx}^{\rm cf}$, a complete resolution of the issue necessarily requires the study of the effects of quenched disorder, present in any real system, to which we turn next.
 
\section{Composite fermion mean-field theory with disorder}

The effects of chemical potential disorder can be incorporated as a shift $\mu \rightarrow \mu + V({\bf x})$, where $V({\bf x})$ is a random potential. 
We take $V({\bf x})$ to satisfy 
\begin{equation}
\overline{V({\bf x})} = 0, \ \ \  \overline{V({\bf x})V({\bf x'})}= g e^{-(\bm{x}-\bm{x}')^2/\mathcal{R}^2}, 
 \end{equation}
 where the overline denotes averaging with respect to the disorder distribution and ``$\mathcal{R}$" represents the range of disorder.   
In the composite fermion frame, the disorder has two manifestations.  First, since composite fermions are charged under $A_{\mu}$, they directly couple to the random chemical potential, which, in turn, leads to a spatially varying composite fermion density.  
Second, the flux attachment constraint, $4 \pi \langle f^{\dagger} f (\bm r) \rangle = - \epsilon_{ij} \langle \partial_i a_j (\bm r) \rangle$, implies the composite fermions also experience a random magnetic field $b({\bf x})$ of zero mean due to the spatially-varying composite fermion density.  
Thus, within a mean-field approximation where the fluctuations of the emergent gauge field are ignored, the effect of disorder on the composite fermion metal is equivalent to that of a free fermion gas in the presence of both potential and flux disorder, with a precise local correlation between the two disorder variables:
 \begin{equation}
 \label{slaved}
b({\bf x}) = - 4 \pi \kappa V({\bf x})
 \end{equation}
where $\kappa$ is the local compressibility that relates the local composite fermion density to their local chemical potential. 
Being linear in $V({\bf x})$, Eq.~\eqref{slaved} assumes sufficiently weak chemical potential disorder.
 
\begin{figure}
\includegraphics[width=3in]{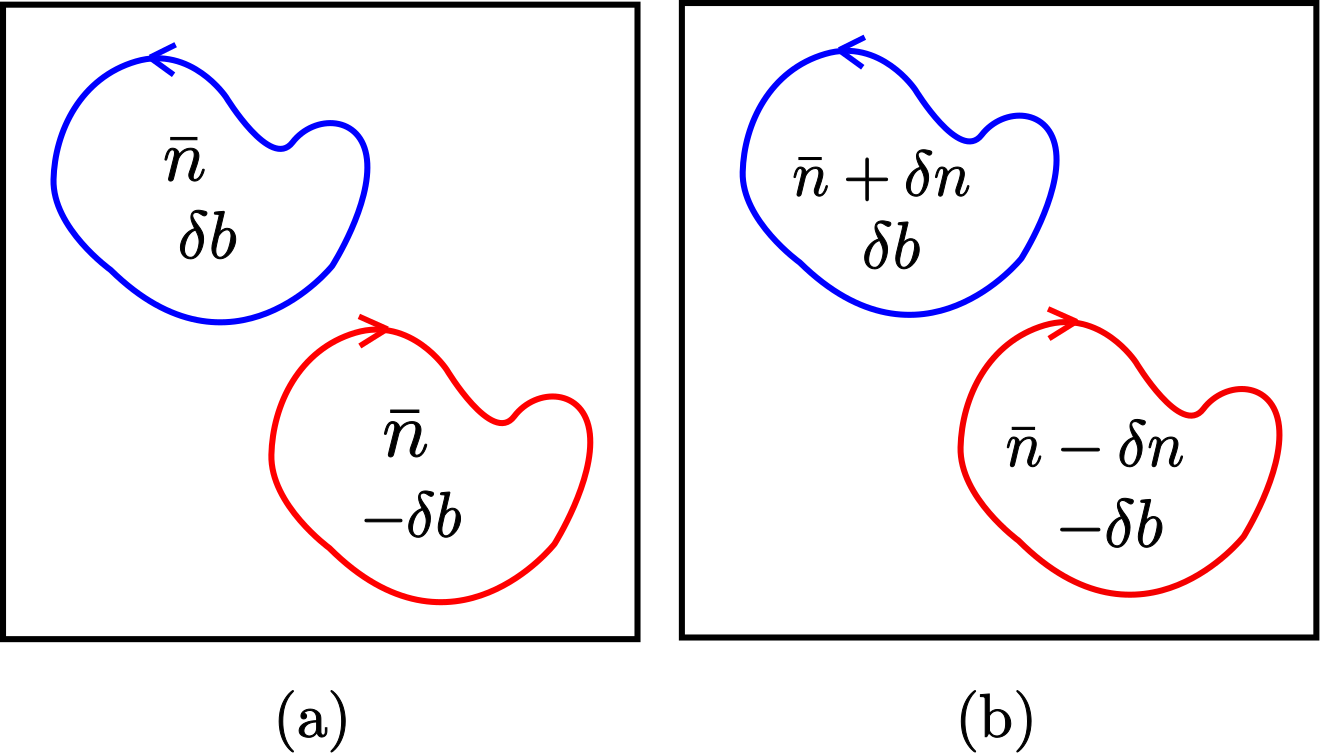}
\caption{(a) A cartoon of a typical spatial configuration with pure flux disorder. 
Such disorder corresponds to an incorrect treatment of chemical potential disorder (as viewed by the electrons) within the HLR theory.
(b) A cartoon of a typical spatial configuration of random composite fermion density and magnetic flux disorder ``slaved" to one
another as in Eq.~\eqref{slaved}.
In both cases, the system has an average density $\bar n$. }
\label{blob}
\end{figure}

In the remainder, we will focus upon the effects of random potential and magnetic disorders -- {\it slaved} to one another as in Eq.~\eqref{slaved} -- on the Hall conductivity of free composite fermions, {\it i.e.}, the mean-field approximation to the HLR composite fermion theory.
When both magnetic and potential disorders are slaved to one another in this way, new features can arise that are not present for either magnetic or potential disorder in isolation (or together, but uncorrelated to one another).  
As is well-known, all states of a free 2D Fermi gas in the presence of a disordered chemical potential are localized.\cite{Anderson1979}  Furthermore, in the presence of purely magnetic disorder with {\it zero} average magnetic field, all states of a free 2D Fermi gas also remain localized.\cite{Hikami80}\footnote{There can be extended states at half-filling if the lattice is bipartite.  
However, this is highly non-generic, and we will ignore this possibility}  
This is because for every region with flux $\delta b$, there is an equivalent contribution from another region with the opposite flux $- \delta b$.  
The average filling fraction then is zero and although time-reversal is broken in each disorder realization,  there is a `statistical' time-reversal symmetry, obtained upon disorder averaging (Fig. \ref{blob}(a)).   

By contrast, when both potential and flux disorders are slaved to one another, statistical time-reversal symmetry is lost: intuitively, a non-zero Hall effect is possible in such a system since a larger density of particles have a counterclockwise cyclotron motion (see Figure \ref{blob} (b)).  As a heuristic estimate, consider two typical regions: one with density $\bar n + \delta n$ and another with density $\bar n - \delta n$ as in Fig. \ref{blob}(b).  Since the filling fractions in these regions have opposite signs due to the slaving of disorder, the Hall conductance from these regions is 
\begin{align}
\sigma_{xy}^{\rm cf} \simeq {1 \over 2 \pi} \nu_{\rm eff} = {\delta n \over \delta b} \approx - {1 \over 4\pi}.
\end{align}
 Such a motion in general can produce a non-quantized Hall effect, which in turn, for free fermions, implies extended states at the Fermi energy.  The only question remains whether or not such cyclotron motion  percolates, leading to observable transport.  To address this question, we study the problem numerically on lattices with slaved potential and flux disorder.
  
 \section{Numerical Analysis}
 \subsection{Continuum model}
Within the mean-field approximation, we treat composite fermions as free nonrelativistic fermions in the presence of slaved disorder. The single-particle Hamiltonian is taken to be
 \begin{equation}
 \label{CF_H_continuum}
 \mathcal{H}_{\rm cf} = \frac{(p_j-a_j)^2}{2m_f} - V(\bm{x}),
 \end{equation}
 where $m_f$ is the composite fermion mass and $a_j$ is the fluctuation of the vector potential that is slaved by Eq.~\eqref{slaved} to the potential disorder $V({\bf x})$.

To first order in the disorder potential, the fluctuation of the composite fermion density $\delta n(\bm{k}) = \kappa (\bm{k}) V(\bm{k})$, where $\delta n(\bm{k})$ represents the Fourier components of composite fermion density modulation due to the Fourier transformed disorder potential $V({\bf k})$. For a 2D free Fermi gas, $\kappa(\bm{k}) = m_f/2\pi$ for $|\bm{k}| < 2k_F$. So, we assume that $\kappa(\bm{k}) = m_f/2\pi$ as long as the Fourier components of potential with $|\bm{k}|>2k_F$ are small. Consequently, Eq.~\eqref{slaved} becomes:
 \begin{equation}
 \epsilon_{ij} \partial_i a_j({\bf x}) = b(\bm{x}) = - 2 m_f V(\bm{x}).
 \end{equation}
 
 We model $V(\bm{x})$ as a superposition of randomly-located impurities each with an individual potential given by:
 \begin{equation}
 V(\bm{x})=\frac{2V_0}{\mathcal{R}\sqrt{\pi n_i}} e^{-2|{\bf x}|^2/\mathcal{R}^2}.
 \end{equation}
where $n_i$ is the number of impurities per unit area and the variance of disorder is given by $V_0^2$. We choose the number of attractive impurities to be equal to the number of repulsive impurities so that  odd moments of the disorder potential vanish when disorder averaged, thereby, ensuring that the disorder is particle-hole symmetric in a statistical sense.
In addition, $\overline{ V(\bm{x}_0)V(\bm{x}_0 + \bm{x})} \propto e^{-|{\bf x}|^2/\mathcal{R}^2}$.
 
 For a numerical calculation, one needs to consider a finite dimensional Hilbert space. First, we place composite fermions on a finite system with size $L\times L$. The calculations are performed by taking plane waves as the basis of the Hilbert space. Next, we choose a cutoff $\Lambda$ on the allowed momenta $|\bm{k}| \leq \Lambda$ in such a way that the results are insensitive to the precise value of the cutoff. 
Because the cutoff breaks gauge invariance, it is important to choose a gauge that minimizes the errors due to the circular cutoff. 
We take
 \begin{equation}
 a_j(\bm{k}) = -ib(\bm{k}) \frac{\epsilon _{ij}k_i}{k^2}.
 \end{equation}

 The Hamiltonian in Eq.~\eqref{CF_H_continuum} is transformed to the plane wave basis and it is diagonalized to obtain the wavefunctions and the corresponding energies. Then we use the Kubo formula to calculate the frequency dependent composite fermion Hall conductivity:
 \begin{equation}
  \sigma_{xy}^{\rm cf}(\omega) = -\frac{i2\pi}{L^2}\sum_{l\neq n} \frac{f(E_n)-f(E_l)}{E_n-E_l} \frac{v^x_{nl} v^y_{ln}}{E_n-E_l + \omega + i \eta},
  \label{Kubo}
 \end{equation}
 where $v^j = (p_j-a_j)/m$, $\mathcal{H}_{\rm cf}\left|n\right\rangle = E_n\left|n\right\rangle$, $v^j_{nl} = \left\langle n\right| v^j \left| l\right\rangle$ and $f(E)$ is the Fermi-Dirac distribution. In addition, $\omega$ is the angular frequency of the probing electric field and $\eta$ is a smearing factor.
 
 A temperature of the order of eigenvalue spacing is included so as to reduce fluctuation effects due to disorder. We then average $\sigma_{xy}^{\rm cf}$ over a number of disorder realizations. The exact number of disorder realizations depends on the size of disorder-induced fluctuations. 
 Typically, these fluctuations are larger for weaker disorder and smaller system sizes.
 \begin{figure}
 \includegraphics[scale=0.75]{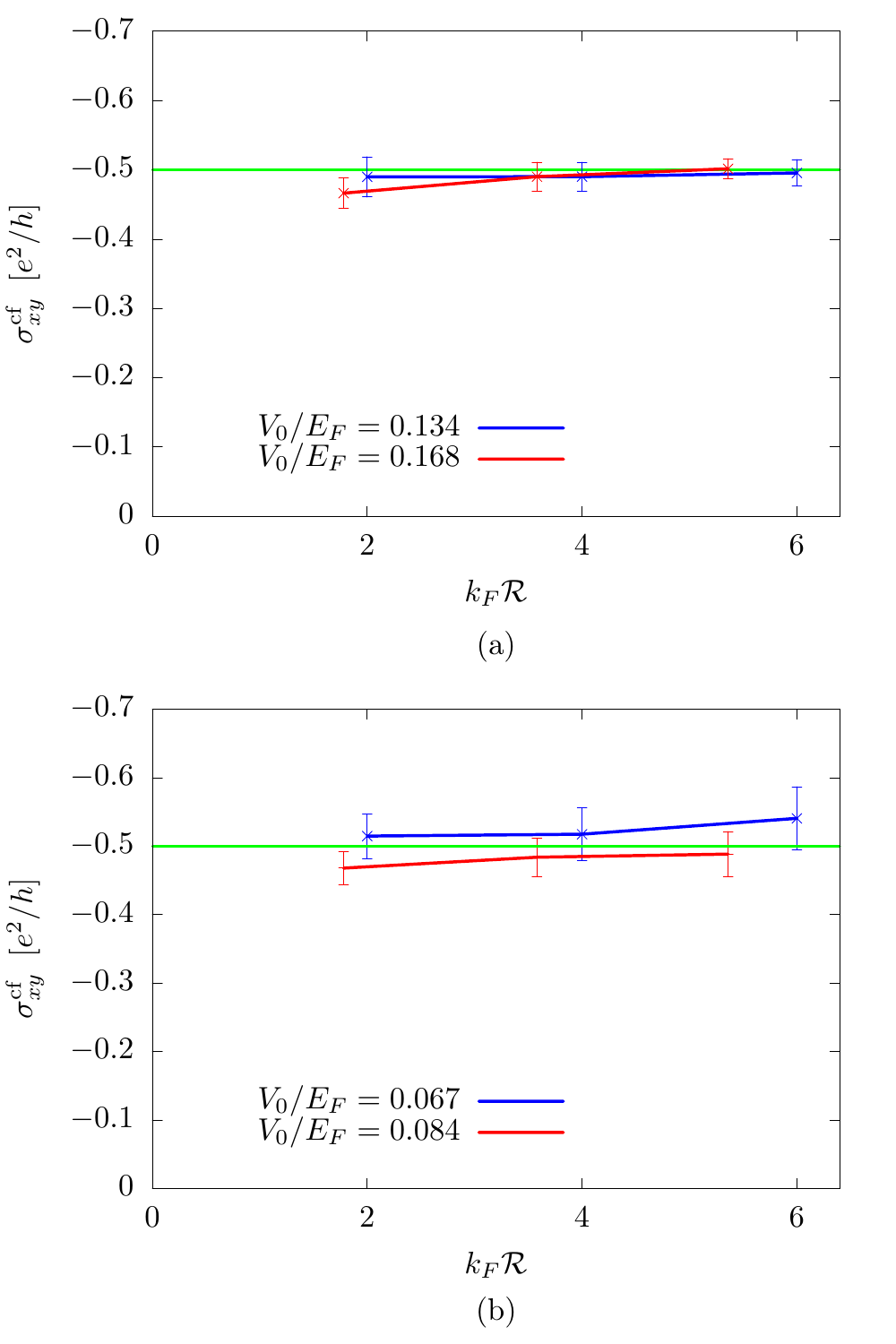}
\caption{dc $\sigma_{xy}^{\rm cf}$ is plotted for two different strengths of disorder in the continuum model defined by Eq.~\eqref{CF_H_continuum} in (a) and (b). The system size $L$ ranges from $50/k_F$ to $67/k_F$ and the cutoff $\Lambda$ ranges from $3k_F$ to $3.35 k_F$. In each of the plots, two different Fermi energies are considered. 
}
\label{continuum_plots}
 \end{figure}
 
 \subsubsection{Continuum results}
 \textbf{dc Hall conductivity:} We plot the computed composite fermion dc Hall conductivity in Fig.~\ref{continuum_plots} as a function of the dimensionless parameter $k_F \mathcal{R}$. We have disorder averaged it over approximately $2\times 10^4$ disorder realizations and set $\eta=0$. It can be seen that the composite fermion Hall conductivity equals $- 1/4\pi$ (up to numerical error) when $k_F \mathcal{R}$ is large. Remarkably, $\sigma_{xy}^{\rm cf}$ maintains its particle-hole symmetric value even when $k_F\mathcal{R} \sim 1$. 
This behavior appears to persist as the disorder strength is varied, indicating some robustness of this result to the strength and range of disorder.

\begin{figure}
 \includegraphics[scale=0.75]{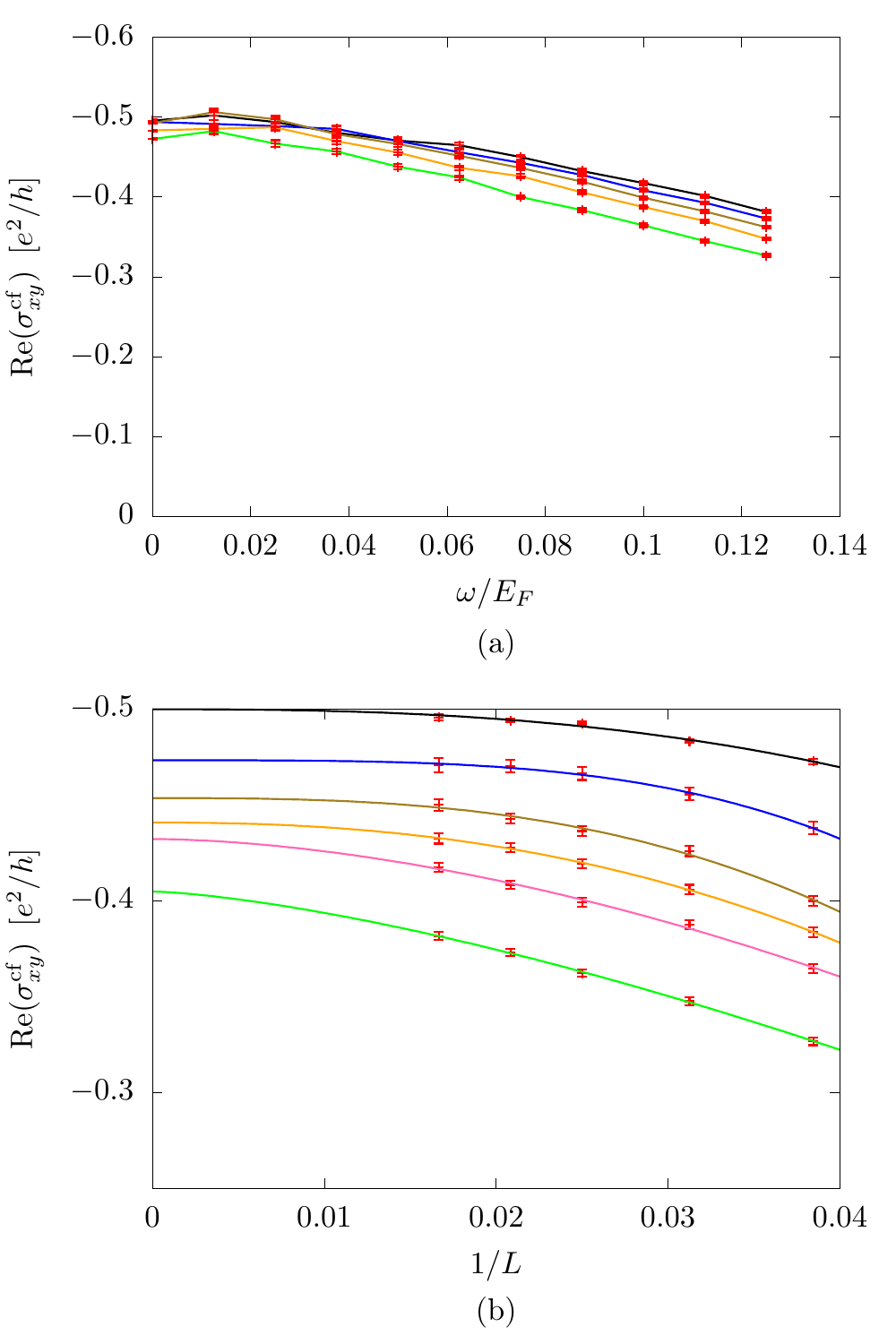}
 \caption{
 (a) $\mathrm{Re}(\sigma_{xy}^{\rm cf})$ vs $\omega/E_F$ for various system sizes, $Lk_F = 54, 43, 36, 29$ and $23$ (from top to bottom). (b) $\mathrm{Re}(\sigma_{xy}^{\rm cf})(\omega)$ vs $1/L$ for $ \omega/E_F = 0,0.05,0.075,0.088,0.1$ and $0.125$ (from top to bottom).
 These frequencies are chosen because their error bars are relatively small.
 The data has been fitted to a power law, i.e. $\mathrm{Re}(\sigma_{xy}(\omega)) = \mathrm{Re}(\sigma_{xy}(\omega))_{L=\infty} + a L^{-\gamma}$.
In both plots, the momentum space cutoff $\Lambda = 3.6 k_F$, the disorder strength $V_0/E_F = 0.19$, and the disorder range $k_F\mathcal{R} = 5.4$. }
 \label{hall_vs_L}
 \end{figure}

\textbf{ac Hall conductivity:} We numerically calculate the ac composite fermion Hall conductivity $\sigma^{\rm cf}_{xy}(\omega)$ for a range of values of frequencies $\omega$ and system sizes $L$ at a fixed Fermi energy $E_F$ and fixed disorder strength. 
Because the real part of $(E_n-E_l + \omega + i\eta)$ in Eq.~\eqref{Kubo} is generally smaller for $\omega > 0$ than when $\omega=0$,
we have used a small smearing factor $\eta=0.1\times 2\pi m_f/L^2$ to reduce fluctuations. 
For the largest system size considered, the disorder average is taken over $10^5$ disorder realizations. 
For smaller system sizes, the number of disorder realizations is larger. The results are plotted in Fig.~\ref{hall_vs_L}. 
We notice that at $\omega>0$, $\sigma_{xy}^{\rm cf} (\omega)$ deviates from $-1/4 \pi$.
 The degree of deviation becomes smaller as system grows in size. However, an extrapolation of the data, using a power law in system size, suggests that the ac Hall conductivity is not particle-hole symmetric in the thermodynamic limit.
 In addition, we find the deviation to be stronger for weaker disorder.
%
 
 \subsection{Tight-binding model}
 We repeat a similar analysis for the dc composite fermion Hall conductivity only for a tight-binding Hamiltonian on a $L\times L$ square lattice with nearest-neighbor hopping:
\begin{equation}
\mathcal{H}_{\rm cf} = t\sum_{\langle ij\rangle} c^\dagger_i c_j e^{ia_{ij}} - \sum_i V_{i}c^\dagger_{i}c_{i}.
\label{eq:CF_H_lattice}
\end{equation}
As in the continuum case, we slave the potential and flux disorders locally. 
However, since magnetic flux lives on plaquettes while the potential lives on lattice sites, we take the flux on a plaquette ($\phi_i$) to be proportional to the average of the potential on its four surrounding vertices $V_{{\rm avg},i}$:
\begin{equation}
\phi_i = -4\pi \kappa \times V_{{\rm avg},i}
\end{equation}
The compressibility on the lattice $\kappa$ is calculated self-consistently for the combined disorder and is taken to be independent of the wavevector.
 
 The potential disorder is modeled in the following way:
 \begin{equation}
 V_i = \frac{V_0}{\mathcal{N}}\sum_{j} f_j\, e^{-2(\bm{x}_i-\bm{x}_j)^2/\mathcal{R}^2}
 \end{equation}
 where $f_i \in [-1/2, 1/2]$ are uncorrelated random numbers. $\mathcal{N}$ is determined by the condition that the variance of the disorder is given by $V_0^2$. Also, $\overline{ V_i V_j} \propto e^{-(\bm{x}_i-\bm{x}_j)^2/\mathcal{R}^2}$.
 \begin{figure}
 \includegraphics[scale=0.75]{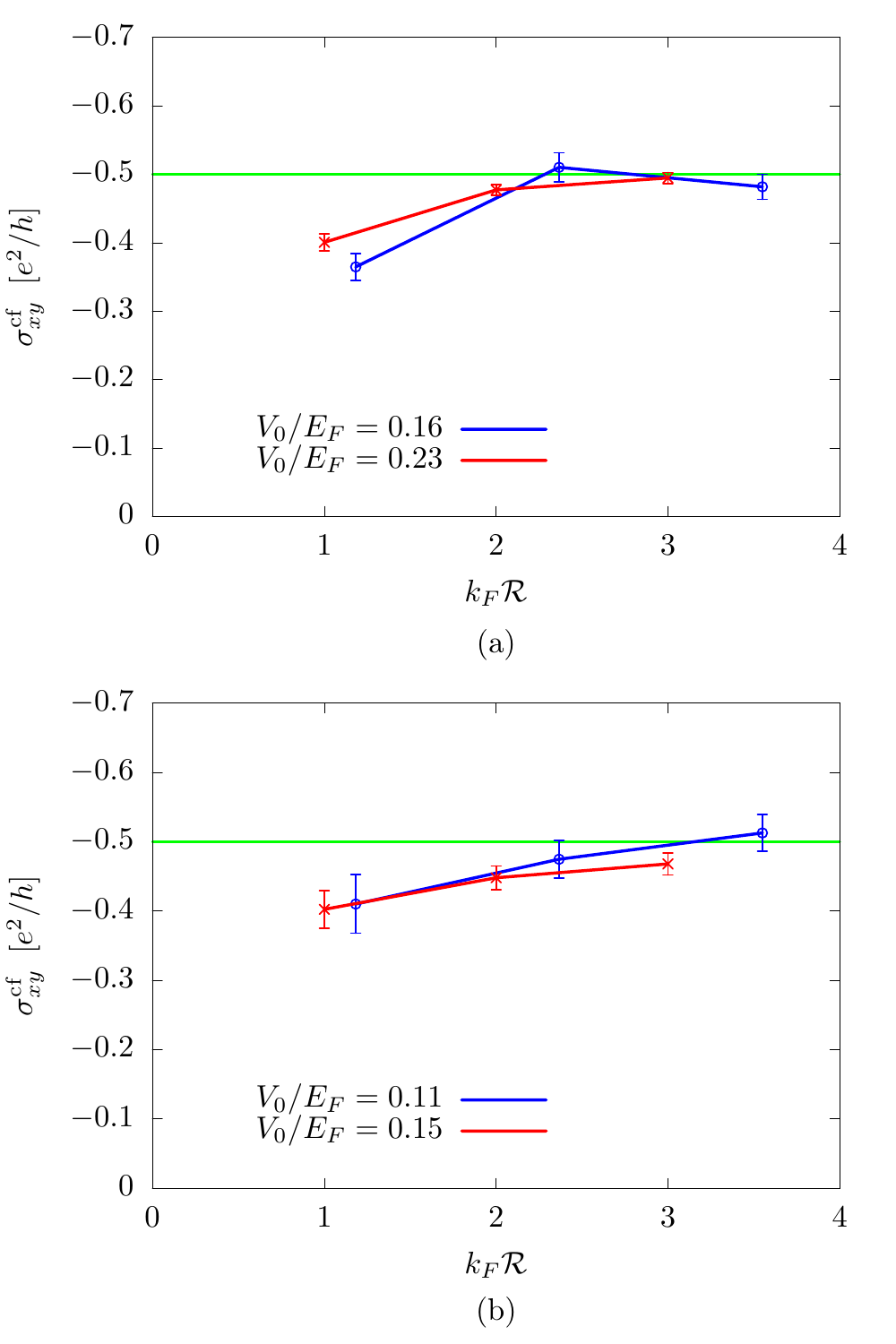}
 \caption{dc $\sigma_{xy}^{\rm cf}$ is plotted for two different strengths of disorder in the tight-binding model defined by Eq.~\eqref{eq:CF_H_lattice} in (a) and (b). The system size $L$ ranges from $60/k_F$ to $70/k_F$. In each of the plots, two different Fermi energies are considered.}
 \label{lattice_plots}
 \end{figure}
 
 The Hamiltonian is diagonalized in real space for the tight-binding model. Then, the dc Hall conductivity of the composite fermions is computed using the Kubo formula given in Eq.~\eqref{Kubo}. 
 In this case, we take the disorder average over approximately $10^4$ realizations.

 \subsubsection{Tight-binding Results}
 The results for tight-binding Hamiltonian are presented in Fig.  \ref{lattice_plots}. We have considered slightly stronger and shorter ranged disorder compared to the continuum case. The composite fermion Hall conductivity shows similar behavior as in  the continuum model for long ranged disorder, i.e., $\sigma^{\rm cf}_{xy} \approx - 1/4\pi$ for large $k_F {\cal R}$. 
 However, deviations from this value appear as $k_F \mathcal{R} \rightarrow 1$.

 \section{Discussion}
 Employing a standard mean-field approximation, which neglects fluctuations of the emergent gauge field, we have shown that the HLR theory can exhibit a particle-hole symmetric composite fermion Hall conductivity equal to $-1/4\pi$ in the presence of quenched disorder.
As stressed in recent works,\cite{2017PhRvX...7c1029W, PhysRevB.98.115105} the key to obtaining this result lies in incorporating the effects of the flux attachment constraint
that relates the local variation in flux to that of the density of composite fermions: $4 \pi \langle n({\bf x}) \rangle = -\langle \epsilon_{ij} \partial_i a_j({\bf x}) \rangle$.  Going beyond leading order in an expansion in $1/{k_F \mathcal R}$, we have shown, using two different microscopic models, that particle-hole symmetric dc response is robustly satisfied to excellent approximation even in the regime $k_F \mathcal R \sim \mathcal O(1)$.  
 In one of these models (continuum), we have also numerically calculated the composite fermion Hall conductivity at finite frequencies.
 We find deviations from particle-hole symmetric response at sufficiently large frequencies compared with the inverse system size, which are reduced as the system size increases.
 
 It is important to consider other tests of an emergent particle-hole symmetry of the $\nu=1/2$ state, both experimentally and theoretically.
 Levin and Son\cite{LevinSon2016} have derived a remarkable linear relation between the electrical Hall conductivity and scalar potential susceptibility that any particle-hole symmetric state must satisfy. 
 Wang and Senthil\cite{PhysRevB.94.245107} have found particle-hole symmetry places a strict constraint on the thermal Hall conductivity of the electronic state.
To date, it is unknown whether experiment or the HLR theory is compatible with either constraint that these works derive.

 \textbf{Acknowledgement}: SR and PK were supported by the DOE Office of Basic Energy Sciences, contract DE-AC02- 76SF00515.
MM was supported in part by the UCR Academic Senate. 
 
 \appendix
 \section{Linear response dictionary between electrons and composite fermions}
 Starting from the composite fermion Lagrangian $\mathcal L_{\rm eff} = \mathcal L_f + \mathcal L_{\rm cs} + \mathcal L_{\rm int}$ with
  \begin{align}
\mathcal L_f  &= f^{\dagger} \left( i \partial_t + A_t + a_t - \frac{1}{2 m_f} \left( i \partial_j + A_j + a_j \right)^2 \right) f, \nonumber \\
\mathcal L_{\rm cs} &= \frac{1}{2} \frac{1}{4 \pi} \epsilon_{\mu \nu \lambda} a_{\mu} \partial_{\nu} a_{\lambda}, \nonumber \\
\mathcal L_{\rm int} &= - \int d^2 x' f^{\dagger}(t, \bm x) f(t, \bm x) U (\bm x - \bm x') f^{\dagger}(t, \bm x')f(t, \bm x'), \cr
\end{align}
we formally integrate out the fermions to obtain
\begin{eqnarray}
\mathcal L_{\rm eff}[A,a]&=& \frac{1}{2} \left( A_{\mu} + a_{\mu} \right) \Pi^{\rm cf}_{\mu \nu} \left( A_{\nu} + a_{\nu} \right) + \frac{1}{2} a_{\mu} \Pi^{\rm cs}_{\mu \nu} a_{\nu}  \nonumber \\
\end{eqnarray}
where   $\Pi^{\rm cf}$ is the gauge interaction mediated by the composite fermions and $ \Pi^{\rm cs}_{\mu \nu} = \frac{1}{4 \pi} \epsilon_{\mu \lambda \nu} \partial_{\lambda}$ represents the statistical interaction due to the Chern-Simons term.  The above  is more a definition of an exact composite fermion response Lagrangian rather than a result of a perturbative calculation.  To read off the electromagnetic response, we integrate out $a_{\mu}$ and define
\begin{align}
{\cal L}_{\rm EM} =& \frac{1}{2} A_{\mu} \Pi^{\rm EM}_{\mu \nu} A_{\nu} 
\end{align}
where
\begin{align}
\label{dictionary}
\Pi^{\rm EM}_{\mu \nu}  =& \Pi^{\rm cf}_{\mu \alpha} \left[ \Pi^{\rm cf} + \Pi^{\rm cs} \right]^{-1}_{\alpha \beta} \Pi^{\rm cs}_{\beta \nu}.
\end{align}
Eq.~\eqref{dictionary} is a dictionary relating conductivities of electrons to those of composite fermions.  Working in the gauge $a_t = A_t = 0$, we can express the above kernels as 
\begin{eqnarray}
\Pi^{\rm EM} &=& \frac{i \omega}{2 \pi} \left( \begin{array}{cc} \sigma_{xx} & \sigma_{xy} \\ - \sigma_{xy} & \sigma_{xx} \end{array} \right) \nonumber \\
\Pi^{\rm cf} &=&  \frac{i \omega}{2 \pi} \left( \begin{array}{cc} \sigma^{\rm cf}_{xx} & \sigma^{\rm cf}_{xy} \\ - \sigma^{\rm cf}_{xy} & \sigma^{\rm cf}_{xx} \end{array} \right) \nonumber \\
\Pi^{\rm cs} &=& \frac{i \omega}{2 \pi} \left( \begin{array}{cc} 0 & 1/2 \\ - 1/2 & 0 \end{array} \right).
\end{eqnarray}
After some elementary algebra, we find that the following relation between resistivity tensor of electrons ($\rho_{ab}$) and composite fermions ($\rho_{ab}^{\rm cf}$):
\begin{eqnarray}
\rho_{ab} &=&  - 4 \pi \epsilon_{ab} + \rho_{ab}^{\rm cf},  \ \ \ 
\epsilon_{ab} = \left( \begin{array}{cc} 0 & 1 \\ - 1 & 0 \end{array} \right).
 \end{eqnarray}
 Inverting the resistivity relation above,  it follows that 
 \begin{align}
 \sigma_{xy} = {1 \over 4 \pi} {4 \pi \sigma^{\rm cf}_{xy} + 16 \pi^2 (\sigma^{\rm cf}_{xx})^2 + 16 \pi^2 (\sigma^{\rm cf}_{xy})^2 \over 1 + 8 \pi \sigma^{\rm cf}_{xy} + 16 \pi^2 (\sigma^{\rm cf}_{xx})^2 + 16 \pi^2 (\sigma^{\rm cf}_{xy})^2}.
 \end{align}
 If $\sigma_{xx}^{\rm cf} \neq 0$, the requirement of particle-hole symmetry, namely that $\sigma_{xy} = 1/4\pi$, implies that 
 \begin{equation}
 \sigma_{xy}^{\rm cf} = - 1/4\pi.
 \end{equation}

\bibliography{bigbib}{}
\bibliographystyle{utphys}

\end{document}